\documentclass[conference]{IEEEtran}

\usepackage{graphicx}
\usepackage[linesnumbered,ruled,vlined]{algorithm2e}
\usepackage{array}
\usepackage{float}
\usepackage{hyperref}
\def\BibTeX{{\rm B\kern-.05em{\sc i\kern-.025em b}\kern-.08em
    T\kern-.1667em\lower.7ex\hbox{E}\kern-.125emX}}
\begin{document}

\title{Popularity Ranking of Database Management Systems}

\author{\IEEEauthorblockN{Aleem Akhtar}
\IEEEauthorblockA{\textit{Dept. of Computer Science} \\
\textit{Virtual University of Pakistan}\\
aleem.akhtar@seecs.edu.pk}
}

\maketitle

\begin{abstract}
Databases are considered to be integral part of modern information systems. Almost every web or mobile application uses some kind of database. Database management systems are considered to be a crucial element from both business and technological standpoint. This paper divides different types of database management systems into two main categories (relational and non-relational) and several sub categories. Ranking of various sub categories for the month of July, 2021 are presented in the form of popularity score calculated and managed by DB-Engines. Popularity trend for each category is also presented to look at the change in popularity since 2013. Complete ranking and trend of top 20 systems has shown that relational models are still most popular systems with Oracle and MySQL being two most popular systems. However, recent trends have shown DBMSs like Time Series and Document Store getting more and more popular with their wide use in IOT technology and BigData, respectively.
\end{abstract}


\section{Introduction}
\label{Introduction}
Databases are considered to be integral part of modern information systems. Almost every web or mobile application uses some kind of database. Databases are ubiquitously used in data centers and to maintain records in educational institutes, medical and healthcare systems, and all kinds of private and government institutions. Database and database management system (DBMS) are two distinct things. Organized collection of data is called database while DBMS on the other hand is a software which interacts with the database and acts as an interface between the user and database. But database term is usually used to refer to both the DBMS and database itself \cite{alzahrani2016evolution}.

The DBMSs with their accompanying data communications systems, has made possible for users from all industries to develop both batch and online applications in a cost effective and timely manner. These database systems has served as base for most of the applications in every government agency and industry. These systems were driving force behind the sale of mainframe computers during the 1970s. Historians and industry analysts consider DBMS a crucial part from both business and technological standpoint \cite{grad2009guest}. Some of the reasons supplied by analysts are:

\begin{itemize}
	\item An efficient and cost effective method to program complex applications is provided by DBMS without rewriting the data retrieval and access functions for each application.
\item A simple and standard way is provided by them to share data among multiple users and multiple applications.
\item Specialized user-oriented languages were created.
\item Standard interfaces are provided by DBMSs for data communication programs so that the development, testing, and maintenance of online transaction processing applications could be done efficiently in terms of time and cost.
\item Databases are easily managed on various sequential and random-access devices without the application programmer to think about difference between them.
\item Portability is provided i.e. enabling users to move applications from one operating system or platform to other with an ease.
\item Companies marketing these systems became largest independent software products companies in the late 1970s.
\item A tremendous amount of hardware was sold by IBM and many independent storage device and terminal companies.

\end{itemize}

Rest of the paper is divided into four sections. The first section provides a brief description on different types of database systems followed by ranking and trends in the second section. The third section presents analysis on ranking and trends. The fourth and last section concludes the paper.

\section{Types of Datase Management Systems}
\label{typesofdbms}
\begin{figure*}[htbp]
	\centerline{\includegraphics [width=0.97\textwidth]{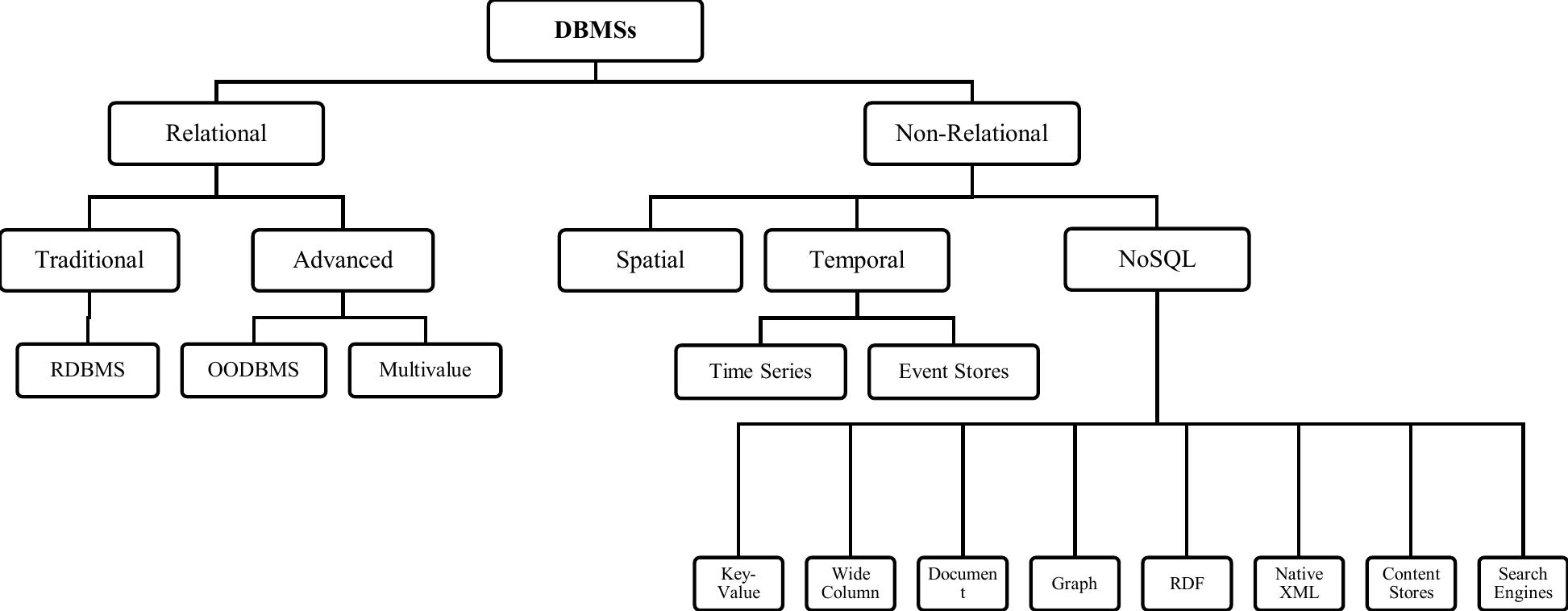}}
	\caption{Types of DBMSs}
	\label{typesofdbmsfig}
\end{figure*}

Database management systems can be generally divided into two main categories: relational database management systems or RDMBS that supports relational data model, and in case it supports other data models are often subsumed as NoSQL systems. However, there are different subcategories of each DBMS and a complete hierarchy is given in Figure \ref{typesofdbmsfig}.

As mentioned earlier, DBMSs can be divided into two broad categories i.e. relational and non-relational. Each category with subcategory is explained below.
\subsection{Relational DBMSs}
Relational database management systems support the relational or table-oriented data model with a pre-defined database schema. Each table/relation has a unique name in the database and fixed number of attributes (columns) with fixed data types. Each row in the table defines a unique record. Normalization is used to generate table schemas during the data modeling process. Relational DBMS allows different types of operations such as classical set operations (intersection, union, and difference), Selection, Projection, and Joins. Some other operations to create, modify, delete table schemas, user management, and transaction controlling are also performed. These basic and advanced operations are performed using some kind of database language, with Structured Query Language (SQL) being well-established standard \cite{sumathi2007fundamentals}.

RDBMS have been most common types of DBMS type since they were first introduced in the early 1980s. Over the years, relational databases have been expanded with advanced non-relational concepts such as non-atomic values (multi-valued), hierarchies, inheritance, and user-defined data types which are sometimes referred to as object-oriented database management systems (OODBMS) \cite{grad2012relational}.

Most popular examples of traditional RDBMS are: Oracle \cite{greenwald2013oracle}\cite{Oracle}, MySQL \cite{widenius2002mysql}\cite{MySQL}, Microsoft SQL Server \cite{mistry2014introducing}\cite{sqlserver}, PostgreSQL \cite{momjian2001postgresql}\cite{PostgreSQL}, and IBM Db2 \cite{karlsson2001ibm}\cite{IBMDb2}. 

\subsubsection{Object Oriented DBMS}
In the 1980s, common use of object-oriented programming languages motivated the development of object-oriented database management systems or simply object databases. Main objective behind introduction of OODBMS was to store the objects and their relationships (inheritance) in the database in a way that relates to their representation in the programming language, without converting or decomposing them \cite{berg2013history}. 

An OODBMS thus follows an object-oriented data model with methods, properties, and classes (objects schema). An object is always managed as a whole. In other words, insertion or reading of object is done in one atomic operation which in relational model will take multiple tables to store that object and complex joins to retrieve it. To perform these operations, a query language similar to SQL is used for manipulation of objects in OODBMS.

Most popular examples of OODBMS are InterSystems Caché \cite{InterSystems}, Db4o \cite{Db4o}, InterSystems IRIS \cite{InterSystemsIRIS}, ObjectStore \cite{ObjectStore}, and Actian NoSQL Database \cite{ActianNoSQL}.

\subsubsection{Multivalue DBMS}
Multivalue DBMS are similar to traditional relational systems and store data in tables. However, major difference between multivalue DBMS and traditional RDBMS is that they have flexibility of storing multiple (non-atomic) values to one attribute. These are often called non-first normal form or NF2 systems as storing non-atomic values contradicts the condition for first normal form. 

Most popular examples of multivalue DBMS are Adabas \cite{Adabas}, UniData,UniVerse \cite{UniDataUniVerse}, jBASE \cite{jBase}, Model 204 \cite{Model204}, and D3 \cite{D3}.

\subsection{Non-Relational DBMS}
Non-relational DBMS are further divided into three subcategories. Each of them is briefly explained.

\subsubsection{NoSQL}
NoSQL database systems do not use a relational data model like RDBMS and generally have no SQL interface. NoSQL databases have been in existence for many years but the term NoSQL was first introduced in 2009 when many new systems were developed in order to cope with the new requirements for DBMS at that time e.g. scalability, Big Data, and fault tolerance \cite{padhy2011rdbms}. Big Data has been at the heart of development of these types of databases. Apache Science Foundation has played one of the most important role in Big Data projects \cite{akhtar2020role}. 

NoSQL systems are a heterogeneous group of very different database systems. Therefore every effort of classification fails in classifying one or another system. However, the following categories are well accepted:
\begin{itemize}
	\item Search Engines
	\item Content Stores
	\item Native XML DBMS
	\item RDF Stores
	\item Graph DBMS
	\item Document Stores
	\item Wide Column Stores
	\item Key-Value Stores
\end{itemize}

Purpose of the paper is to look at the trends of database systems, therefore explanation of each system is skipped.

\subsubsection{Spatial DBMS}
Spatial DBMS is different type of DBMS that can efficiently store, query, and manipulate spatial data. Objects in geometric space such as polygons and points are represented by spatial data. Dedicated data types and spatial indices are provided by spatial DBMSs to optimize the storing and access of spatial data \cite{shekhar2007spatial}. 

Spatial DBMSs provide features of intersecting or merging objects, computing distances, and calculating properties of objects such as areas. Geospatial data are an important subset of spatial data that deals with the locations on surface of Earth.  Geographic Information Systems (GIS) are able to work with geospatial data \cite{rigaux2001spatial}. In some cases, spatial data is combined with temporal data to form spatio-temporal data that offers more dimensions to store and manipulate data. 

Most popular examples of Spatial DBMS are PostGIS \cite{PostGIS}, SpatiaLite \cite{SpatiaLite}, and GeoMesa \cite{GeoMesa}.

\subsubsection{Temporal DBMS}
Temporal DBMS deals with the data related to timestamps or events. Temporal DBMS are classified into two categories i.e. Time Series DBMS and Event Stores.

A \textit{Time Series DBMS} is a database management system that is optimized for handling time series data: each entry is associated with a timestamp \cite{namiot2015time}. For example, time series data may be produced by smart meters, sensors, or RFIDs in the IoT.  Time Series DBMS are aimed to efficiently gather, save and query various time series with high transaction volumes. While time series data can be managed with other categories of DBMS i.e. key-value stores or relational systems, specialized systems are often required to handle specific challenges. Most popular DBMSs in this subcategory are: InfluxDB \cite{InfluxDB}, Kdb+ \cite{Kdb}, and Prometheus \cite{Prometheus}.

\textit{Event stores} are database management systems that implement the event sourcing concepts. All state changing events for an object are preserved by these systems with a timestamp. To inferred current state of an object, all the events from time 0 to current time are replayed. On the other hand, other types of DBMS lose the history of previous states if not modelled explicitly. IBM Db2 is most popular system in this category \cite{garcia2020db2}. 

\section{Ranking and Trends}
\label{rankingsandtrends}

\subsection{Method to calculate score}
DB-Engines website is used as source for all the ranking and trends screenshots. DB-Engines uses a custom formula to calculate a normalized score to rank popular database management systems \cite{DBEngines2021Methods}. Popularity ranking does not measure the DBMSs use within IT systems or the total number of installations on the systems. However, popularity score relate to broad use of a certain system. Following parameters are used by the DB-Engines ranking system to calculate popularity score:

\begin{itemize}
	\item Number of job offers in which system is mentioned on Indeed and Simply Hired websites.
	\item Frequency of the searches in Google Trends
	\item Number of search results in the Google and Bing search engines for particular system
	\item Frequency of the technical discussion on Stack Overflow and DBA Stack Exchange
	\item Number of profiles on LinkedIn with the mention of certain system
	\item Number of tweets where certain system is mentioned using Hashtag
\end{itemize}

\subsection{Ranking and Trends by Type}
In this section, we present ranking of top 5 most popular database management systems by each type for July, 2021. Trends display popularity of these systems from November 2012 to July, 2021. Logarithmic scale is used to display popularity score. 

\subsubsection{Relational DBMS}
Table \ref{rrdbms} presents ranking of top 5 relational DBMS for the month of July, 2021 \cite{rrdbmsDBEngines2021}. Oracle being the most popular system followed by MySQL at second rank. Score for these two systems is very close to each other showing how popular these two systems are. 

\begin{table}[h!]
	\centering
	\caption{Ranking of Relational DBMS}
	\label{rrdbms}
	\begin{tabular}{|c|c|c|}
		\hline 
		\textbf{Rank} & \textbf{DBMS} & \textbf{Score} \\ 
		\hline 
		1 & Oracle  & 1262.66 \\ 
		\hline
		2 & MySQL  & 1228.38 \\ 
		\hline
		3 & Microsoft SQL Server  & 981.95 \\ 
		\hline
		4 & PostgreSQL  & 577.15 \\ 
		\hline
		5 & IBM Db2 & 165.15 \\ 
		\hline
	\end{tabular}
\end{table} 

\begin{figure}[htbp]
	\centerline{\includegraphics [width=0.47\textwidth]{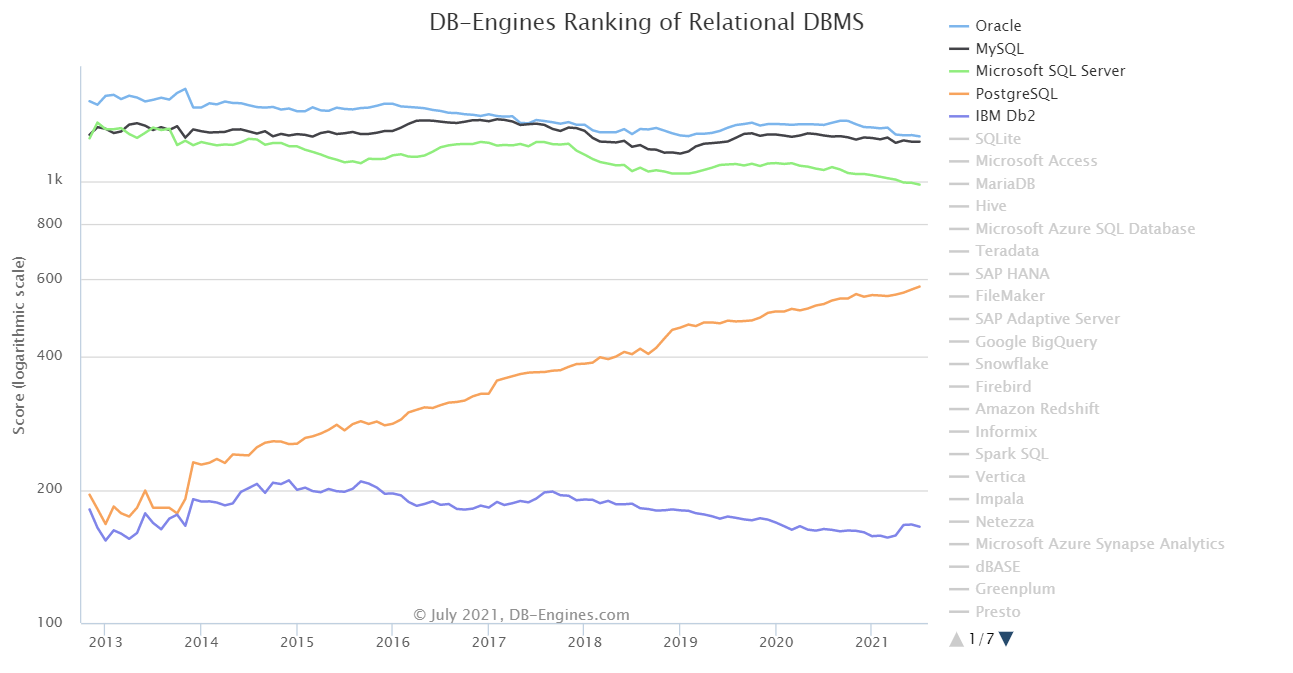}}
	\caption{Trend of Relational DBMS \cite{trdbmsDBEngines2021}}
	\label{trdbms}
\end{figure}

Figure \ref{trdbms} shows popularity trend of top five relational database management systems. Oracle, MySQL, and Microsoft SQL Server have been competing for top spot since 2013. Popularity of PostgreSQL has been on constant rise and is getting closer to the top systems.

\subsubsection{Object Oriented DBMS}
Table \ref{roodbms} presents ranking of top 5 Object Oriented DBMS for the month of July, 2021 \cite{roodbmsDBEngines2021}. A very low score of these systems show that they are still not very popular as other relational systems and it might take a long time to be globally adopted by industry and academia. 

\begin{table}[h!]
	\centering
	\caption{Ranking of Object Oriented DBMS}
	\label{roodbms}
	\begin{tabular}{|c|c|c|}
		\hline 
		\textbf{Rank} & \textbf{DBMS} & \textbf{Score} \\ 
		\hline 
		1 & InterSystems Caché & 2.86 \\ 
		\hline
		2 & Db4o & 1.71 \\ 
		\hline
		3 & InterSystems IRIS  & 1.7 \\ 
		\hline
		4 & ObjectStore & 1.48 \\ 
		\hline
		5 & Actian NoSQL Database & 1.4 \\ 
		\hline

	\end{tabular}
\end{table} 

\begin{figure}[htbp]
	\centerline{\includegraphics [width=0.47\textwidth]{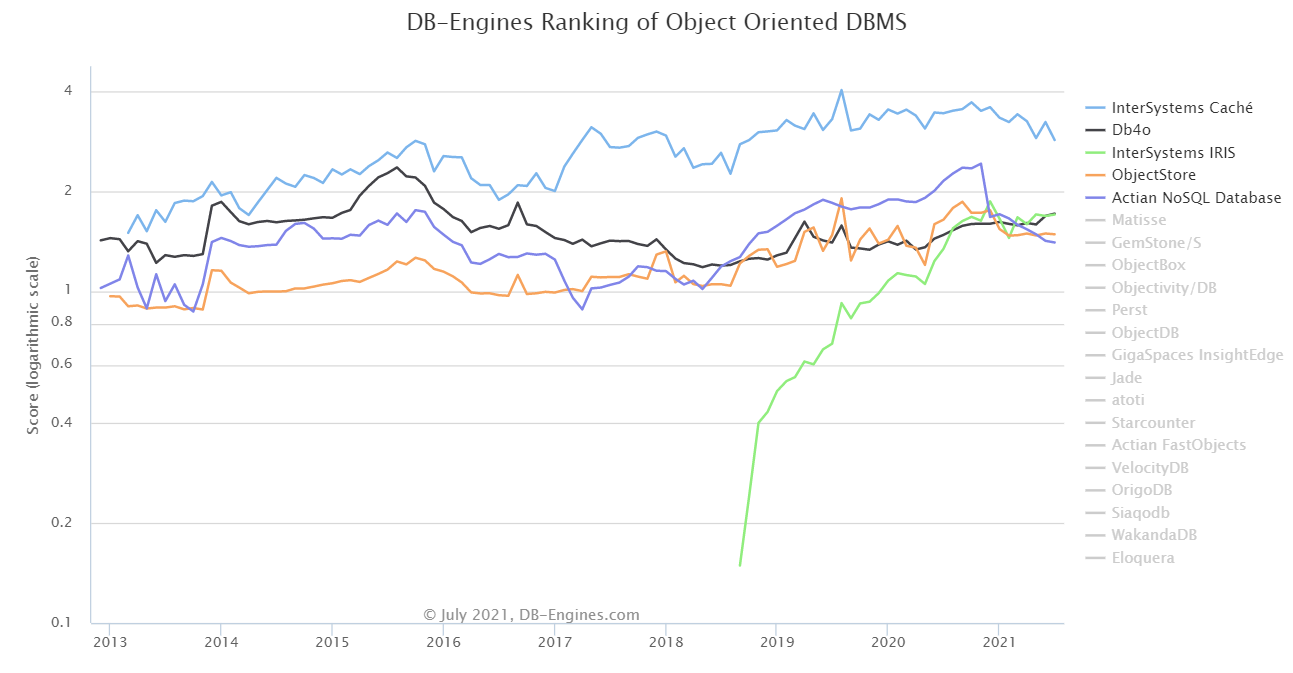}}
	\caption{Trend of Object Oriented DBMS \cite{toodbmsDBEngines2021}}
	\label{toodbms}
\end{figure}

Figure \ref{toodbms} shows popularity trend of top five Object Oriented database management systems. It can be clearly see in the graph that all systems in this category have average popularity score between 1 and 2, and nothing can be said sure about which system is more popular than other.

\subsubsection{Multivalue DBMS}
Table \ref{rmvdbms} presents ranking of top 5 Multivalue DBMS for the month of July, 2021 \cite{rmvDBEngines2021}. Similar to OODBMSs, a very low score of these systems show that they are also not very popular as other relational systems and it can be said that they are still in developmental phases. 

\begin{table}[h!]
	\centering
	\caption{Ranking of Multivalue DBMS}
	\label{rmvdbms}
	\begin{tabular}{|c|c|c|}
		\hline 
		\textbf{Rank} & \textbf{DBMS} & \textbf{Score} \\ 
		\hline 
		1 & Adabas & 4.46 \\ 
		\hline
		2 & UniData,UniVerse & 3.87 \\ 
		\hline
		3 & jBASE & 1.82 \\ 
		\hline
		4 & Model 204 & 1.24 \\ 
		\hline
		5 & D3 & 1.23 \\ 
		\hline
	\end{tabular}
\end{table} 

\begin{figure}[htbp]
	\centerline{\includegraphics [width=0.47\textwidth]{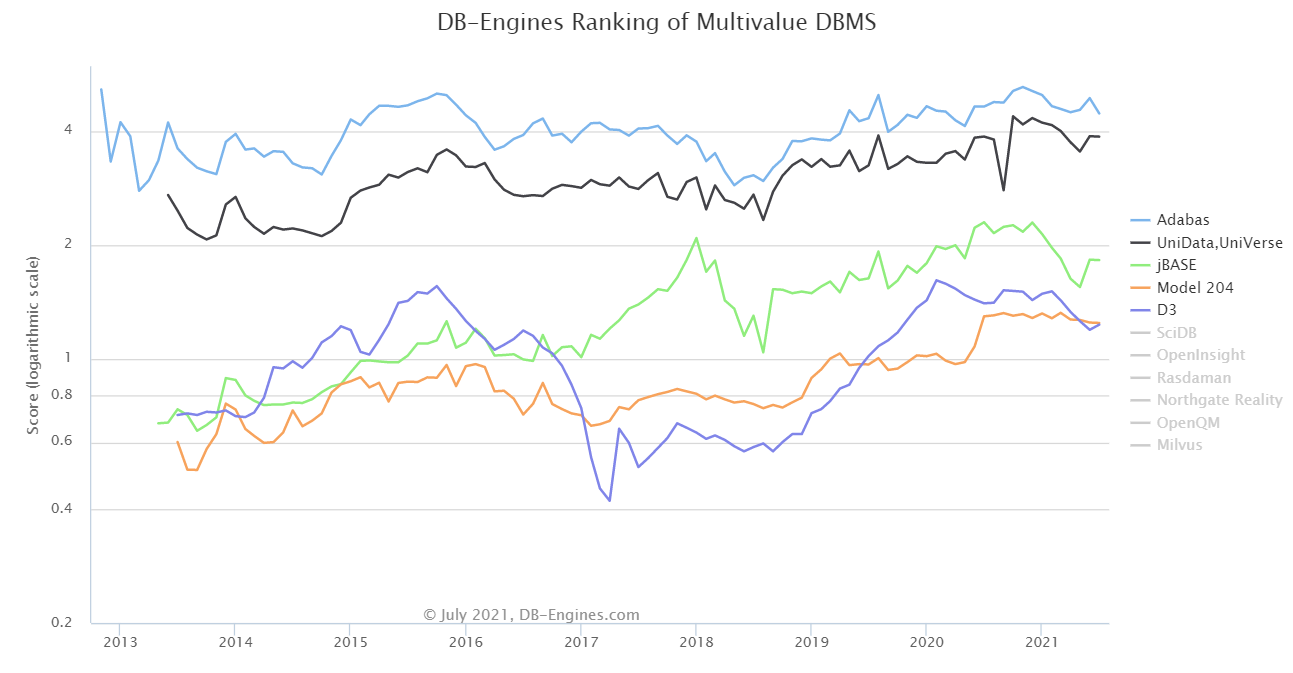}}
	\caption{Trend of Multivalue DBMS \cite{tmvDBEngines2021}}
	\label{tmvdbms}
\end{figure}

Figure \ref{tmvdbms} shows popularity trend of top five Multivalue database management systems. It can be clearly see in the graph that all systems in this category have average popularity score between 1 and 2, and nothing can be said for sure about which system is more popular than other.

\subsubsection{Key-Value Stores}
Table \ref{rkvs} presents ranking of top 5 Key-Value Store DBMS for the month of July, 2021 \cite{rkvsDBEngines2021}. Redis is most popular system from this category with popularity score being more than double compared to second spot Amazon DynamoDB. Score for last three systems is very close to each other but difference between Redis and other systems is very high at this time. 

\begin{table}[h!]
	\centering
	\caption{Ranking of Key-Value Stores DBMS}
	\label{rkvs}
	\begin{tabular}{|c|c|c|}
		\hline 
		\textbf{Rank} & \textbf{DBMS} & \textbf{Score} \\ 
		\hline 
		1 & Redis  & 168.31 \\ 
		\hline
		2 & Amazon DynamoDB  & 75.2 \\ 
		\hline
		3 & Microsoft Azure Cosmos DB  & 36.7 \\ 
		\hline
		4 & Memcached & 25.34 \\ 
		\hline
		5 & etcd & 10.1 \\ 
		\hline

	\end{tabular}
\end{table} 

\begin{figure}[htbp]
	\centerline{\includegraphics [width=0.47\textwidth]{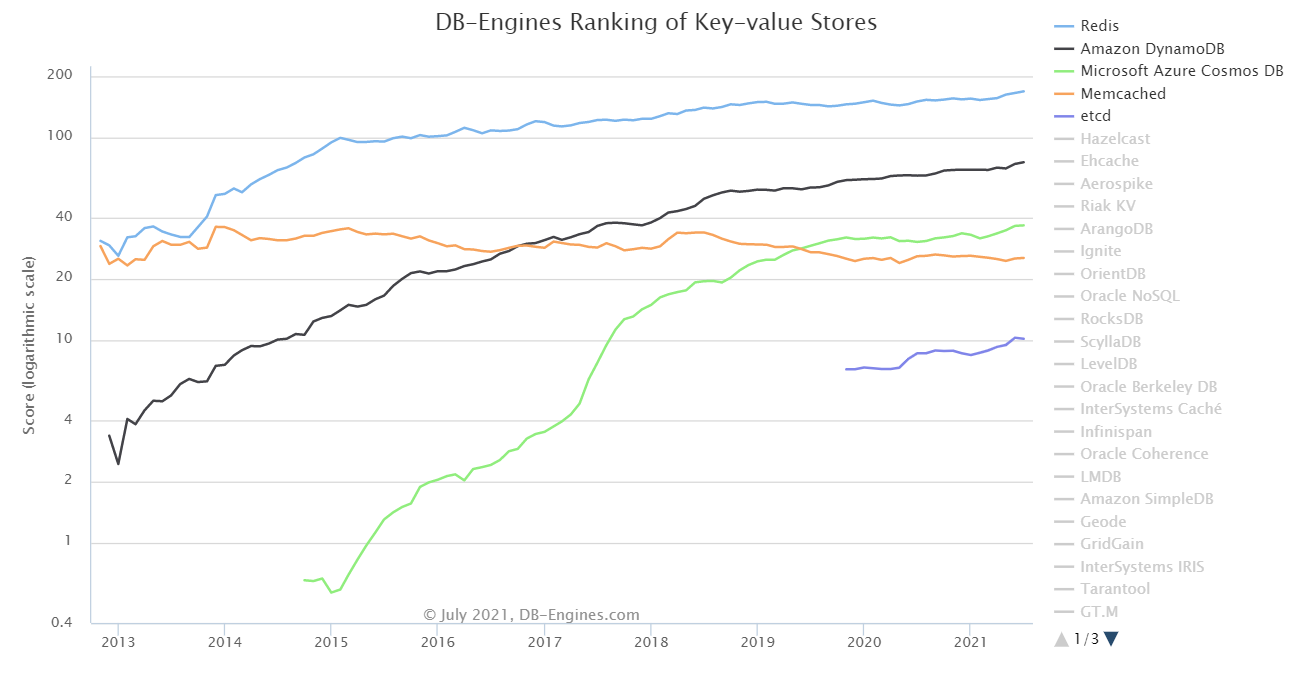}}
	\caption{Trend of Key-Value Stores DBMS \cite{tkvsDBEngines2021}}
	\label{tkvs}
\end{figure}

Figure \ref{tkvs} shows popularity trend of top five Key-Value Store database management systems. Even though Redis has been on the top of popularity chart since 2013 but systems from the Amazon and Microsoft have been on constant rise while on the other hand Memcached is slowly decreasing in popularity.

\subsubsection{Document Store}

Table \ref{rds} presents ranking of top 5 Document Store DBMS for the month of July, 2021 \cite{rdsDBEngines2021}. MongoDB is by far the most popular Document Store DBMS with the popularity score of nearly 500. Amazon DynamoDB and Microsoft Azure Cosmos DB which are also Key-Value Store systems as well come at second and third place respectively. 

\begin{table}[h!]
	\centering
	\caption{Ranking of Document Store DBMS}
	\label{rds}
	\begin{tabular}{|c|c|c|}
		\hline 
		\textbf{Rank} & \textbf{DBMS} & \textbf{Score} \\ 
		\hline 
		1 & MongoDB  & 496.16 \\ 
		\hline
		2 & Amazon DynamoDB  & 75.2 \\ 
		\hline
		3 & Microsoft Azure Cosmos DB  & 36.7 \\ 
		\hline
		4 & Couchbase  & 28.46 \\ 
		\hline
		5 & Firebase Realtime Database & 17.23 \\ 
		\hline

	\end{tabular}
\end{table} 

\begin{figure}[htbp]
	\centerline{\includegraphics [width=0.47\textwidth]{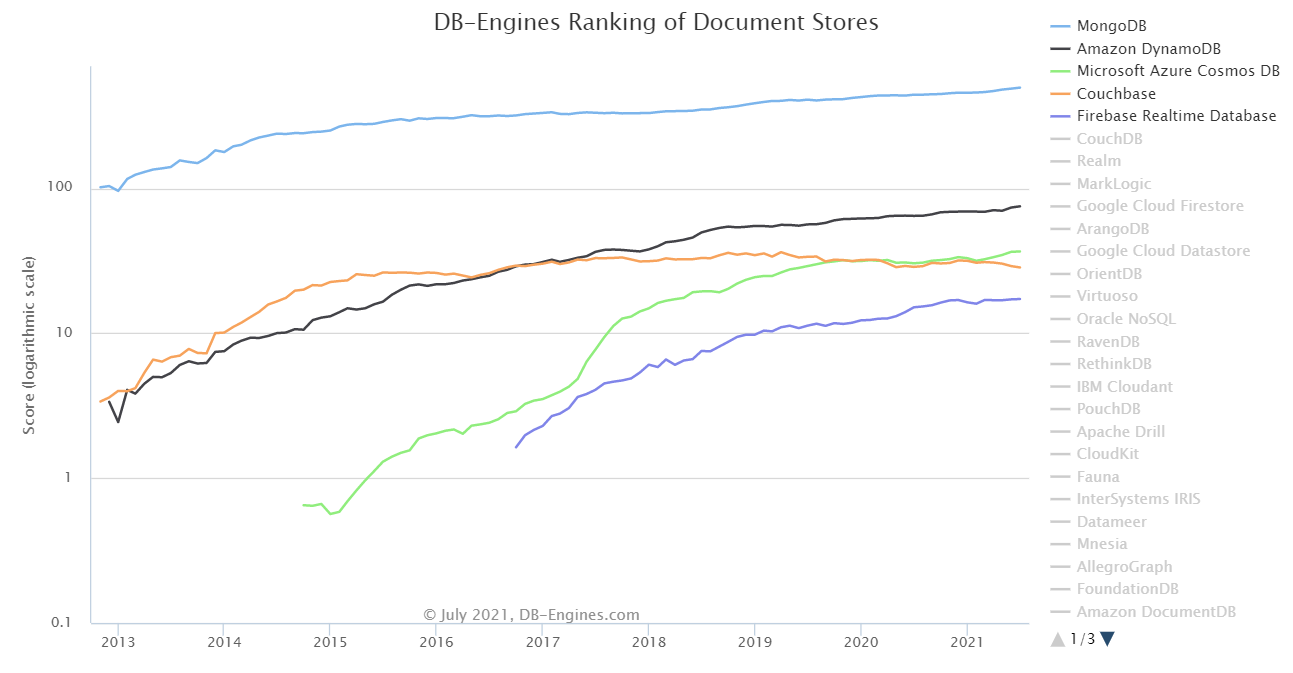}}
	\caption{Trend of Document Store DBMS \cite{tdsDBEngines2021}}
	\label{tds}
\end{figure}

Figure \ref{tds} shows popularity trend of top five Document Store database management systems. MongoDB has been on constant rise in popularity since its introduction but other Document Store DBs have started to gain popularity, Firebase Realtime Database by Google is one of them.

\subsubsection{Wide Column Store}

Table \ref{rwcs} presents ranking of top 5 Wide Column Store DBMS for the month of July, 2021 \cite{rwcsDBEngines2021}. Cassandra being the most popular system followed by HBase at second rank. Score of Cassandra is higher than next four systems combined which shows how popular it is for this category.

\begin{table}[h!]
	\centering
	\caption{Ranking of Wide Column Store DBMS}
	\label{rwcs}
	\begin{tabular}{|c|c|c|}
		\hline 
		\textbf{Rank} & \textbf{DBMS} & \textbf{Score} \\ 
		\hline 
		1 & Cassandra  & 114 \\ 
		\hline
		2 & HBase  & 44.07 \\ 
		\hline
		3 & Microsoft Azure Cosmos DB  & 36.7 \\ 
		\hline
		4 & Datastax Enterprise  & 7.52 \\ 
		\hline
		5 & Microsoft Azure Table Storage & 5.09 \\ 
		\hline
	\end{tabular}
\end{table} 

\begin{figure}[htbp]
	\centerline{\includegraphics [width=0.47\textwidth]{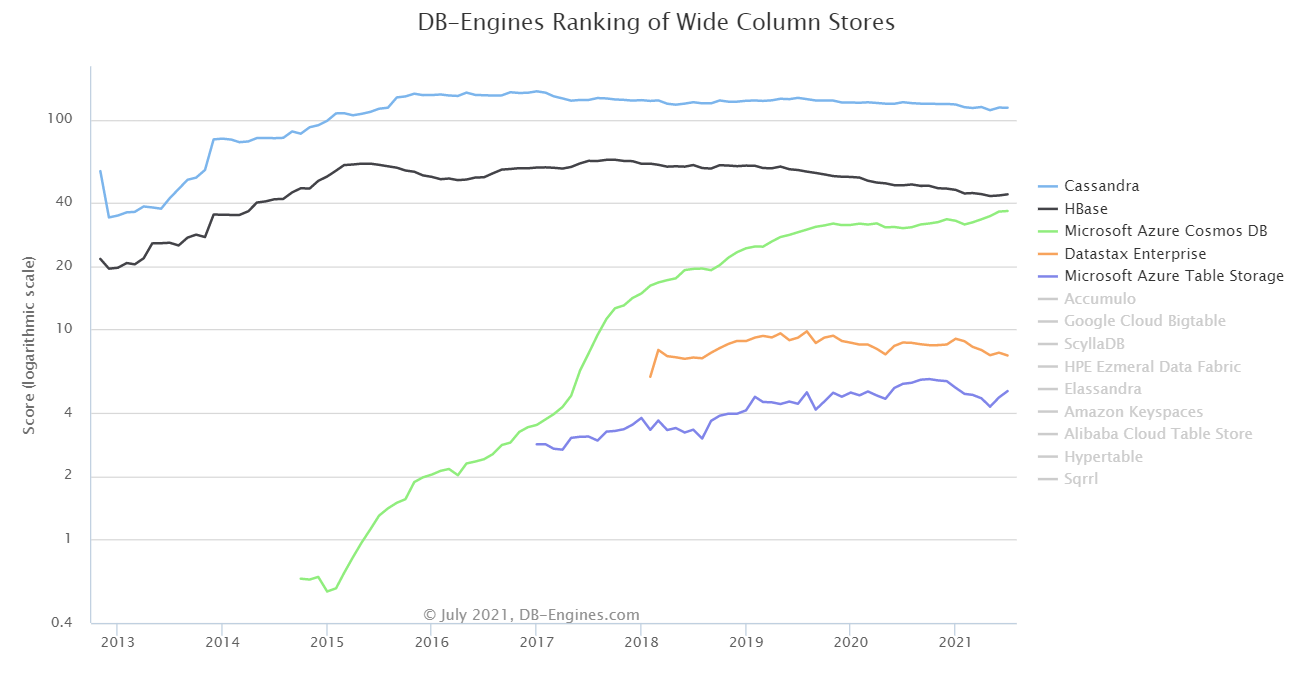}}
	\caption{Trend of Wide Column Store DBMS \cite{twcsDBEngines2021}}
	\label{twcs}
\end{figure}

Figure \ref{twcs} shows popularity trend of top five Wide Column Store database management systems. Oracle and HBase have been competing for top spot since 2013. Popularity of Microsoft Azure Cosmos DB has been on constant rise since its introduction in 2015 and is getting closer to the top systems.

\subsubsection{Graph DBMS}

Table \ref{rgdbms} presents ranking of top 5 Graph DBMS for the month of July, 2021 \cite{rgdbmsDBEngines2021}. Neo4j is most popular system from this category with popularity score being higher than next four systems combined. Score for last three systems is very close to each other but difference between Neo4j and other systems is very high at this time. 

\begin{table}[h!]
	\centering
	\caption{Ranking of Graph DBMS}
	\label{rgdbms}
	\begin{tabular}{|c|c|c|}
		\hline 
		\textbf{Rank} & \textbf{DBMS} & \textbf{Score} \\ 
		\hline 
		1 & Neo4j  & 57.16 \\ 
		\hline
		2 & Microsoft Azure Cosmos DB  & 36.7 \\ 
		\hline
		3 & ArangoDB  & 4.73 \\ 
		\hline
		4 & OrientDB & 4.16 \\ 
		\hline
		5 & Virtuoso  & 4.01 \\ 
		\hline
	\end{tabular}
\end{table} 

\begin{figure}[htbp]
	\centerline{\includegraphics [width=0.47\textwidth]{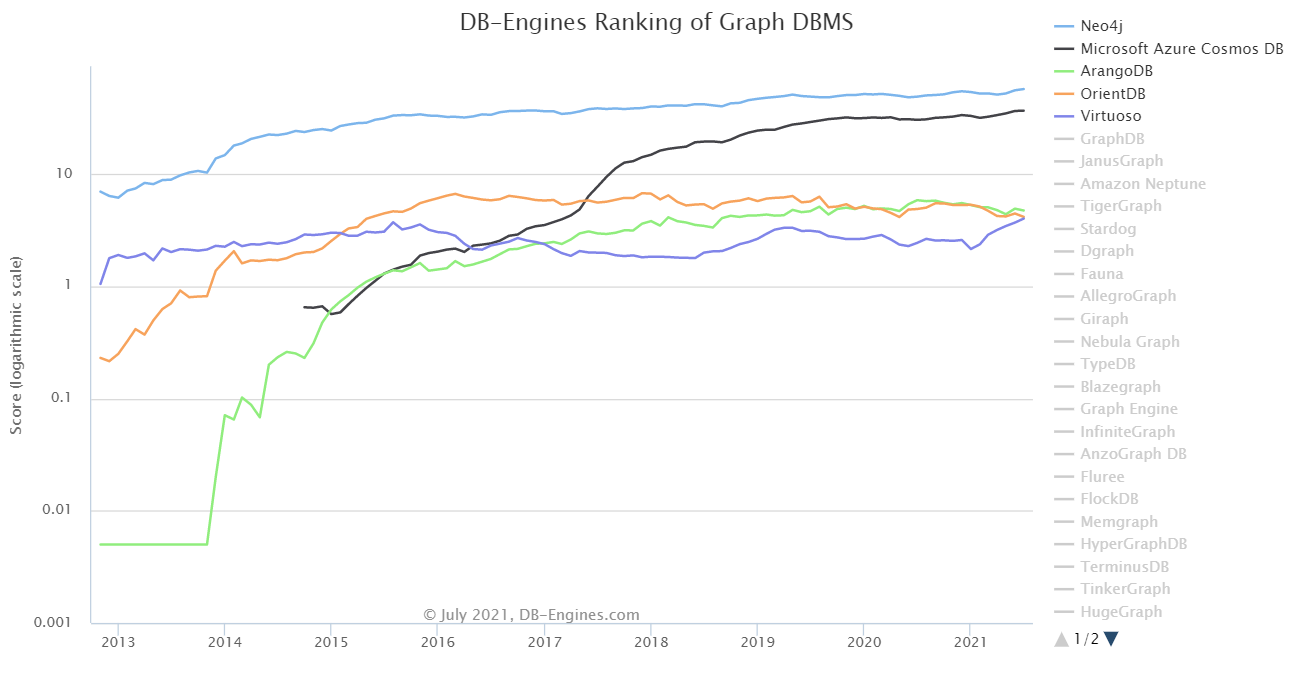}}
	\caption{Trend of Graph DBMS \cite{tgdbmsDBEngines2021}}
	\label{tgdbms}
\end{figure}

Figure \ref{tgdbms} shows popularity trend of top five Graph database management systems. Even though Neo4j has been on the top of popularity chart since 2013 but system from the Microsoft has been on constant rise and getting closer to the most popular system in this category while on the other hand all other systems are slowly decreasing in popularity.

\subsubsection{Search Engines}

Table \ref{rse} presents ranking of top 5 Search Engine DBMS for the month of July, 2021 \cite{rseDBEngines2021}. ElasticSearch is by far the most popular Document Store DBMS with the popularity score of nearly 156. Splunk and Solr come at second and third place respectively followed by MarkLogic and Sphnix at last two spots.

\begin{figure}[htbp]
	\centerline{\includegraphics [width=0.47\textwidth]{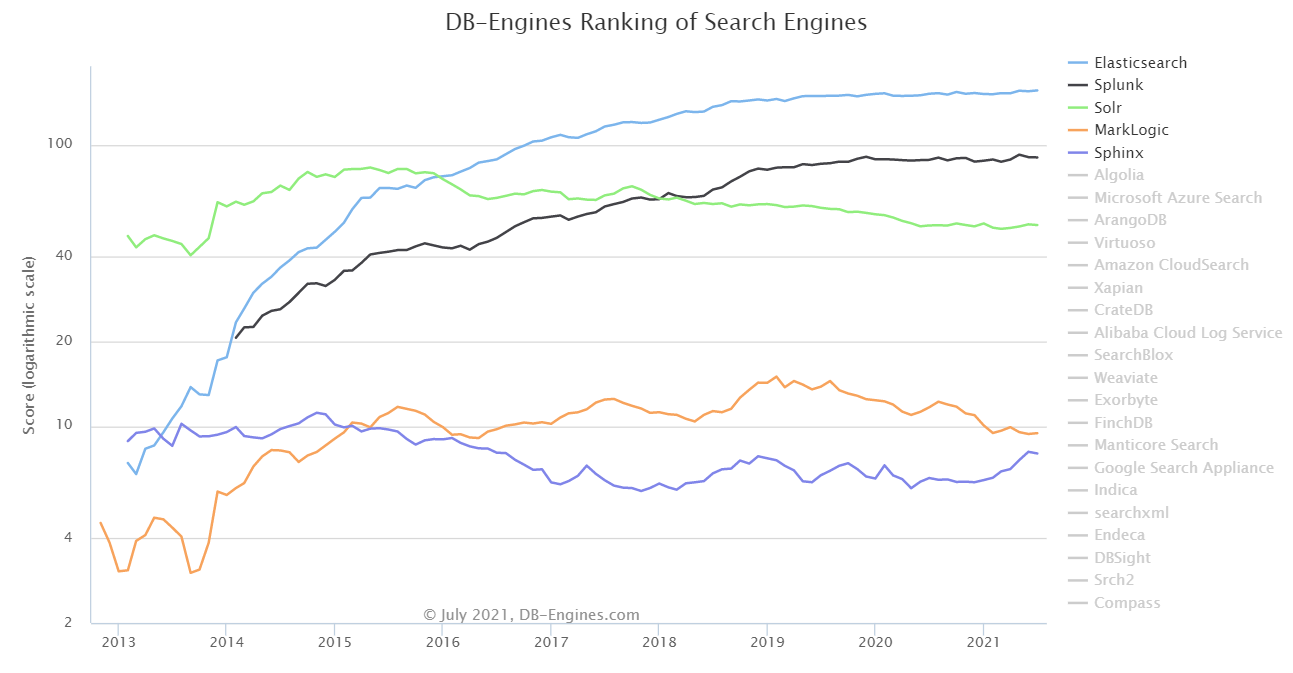}}
	\caption{Trend of Search Engines DBMS \cite{tseDBEngines2021}}
	\label{tse}
\end{figure}

Figure \ref{tse} shows popularity trend of top five Search Engine database management systems. Solr used to be most popular system in this category but it was overtaken by ElasticSearch at the start of 2016. Increase in popularity of Splunk made it second most popular system in January 2018.

\begin{table}[h!]
	\centering
	\caption{Ranking of Search Engines DBMS}
	\label{rse}
	\begin{tabular}{|c|c|c|}
		\hline 
		\textbf{Rank} & \textbf{DBMS} & \textbf{Score} \\ 
		\hline 
		1 & Elasticsearch & 155.76 \\ 
		\hline
		2 & Splunk & 90.05 \\ 
		\hline
		3 & Solr & 51.79 \\ 
		\hline
		4 & MarkLogic  & 9.45 \\ 
		\hline
		5 & Sphinx & 8 \\ 
		\hline
	\end{tabular}
\end{table} 


\subsubsection{Ranking of Top 20}

Table \ref{rcomplete} presents complete ranking of top 20 database management systems for July, 2021 \cite{rankingDBEngines2021}. It can be seen that Oracle and MySQL are top 2 ranked systems with the score of above 1200. Microsoft SQL Server is ranked third followed by PostgreSQL at fourth. All of these top 4 are relational systems which shows the importance of this category. In fact, 12 of the top 20 are relational systems. 3 of them are Search Engines, and one each of Key-Value, Document, Wide Column, and Graph. Amazon Dynamo DB is one system which is a multi-model system. 
Popularity score difference between MongoDB at 5 and Redis at 6 shows that top 5 are solid at these positions for foreseeable future.  

\begin{table}[h!]
	\centering
	\caption{Ranking of Top 20 DBMS}
	\label{rcomplete}
	\begin{tabular}{|c|c|c|c|}
		\hline 
		\textbf{Rank} & \textbf{DBMS} & \textbf{Score}  & \textbf{Type}\\ 
		\hline 
		1 & Oracle  & 1262.66 & Relational \\ 
		\hline
		2 & MySQL  & 1228.38 & Relational \\ 
		\hline
		3 & Microsoft SQL Server  & 981.95 & Relational \\ 
		\hline
		4 & PostgreSQL  & 577.15 & Relational \\ 
		\hline
		5 & MongoDB  & 496.16 & Document \\ 
		\hline
		6 & Redis  & 168.31 & Key-value \\ 
		\hline
		7 & IBM Db2 & 165.15 & Relational \\ 
		\hline
		8 & Elasticsearch  & 155.76 & Search engine \\ 
		\hline
		9 & SQLite  & 130.2 & Relational \\ 
		\hline
		10 & Cassandra  & 114 & Wide column \\ 
		\hline
		11 & Microsoft Access & 113.45 & Relational \\ 
		\hline
		12 & MariaDB  & 97.98 & Relational \\ 
		\hline
		13 & Splunk & 90.05 & Search engine \\ 
		\hline
		14 & Hive & 82.68 & Relational \\ 
		\hline
		15 & Microsoft Azure SQL Database & 75.22 & Relational \\ 
		\hline
		16 & Amazon DynamoDB  & 75.2 & Multi-model  \\ 
		\hline
		17 & Teradata & 68.95 & Relational \\ 
		\hline
		18 & Neo4j  & 57.16 & Graph \\ 
		\hline
		19 & SAP HANA  & 53.81 & Relational \\ 
		\hline
		20 & Solr & 51.79 & Search engine \\ 
		\hline
	\end{tabular}
\end{table} 

\begin{figure}[t]
	\centerline{\includegraphics [width=0.47\textwidth]{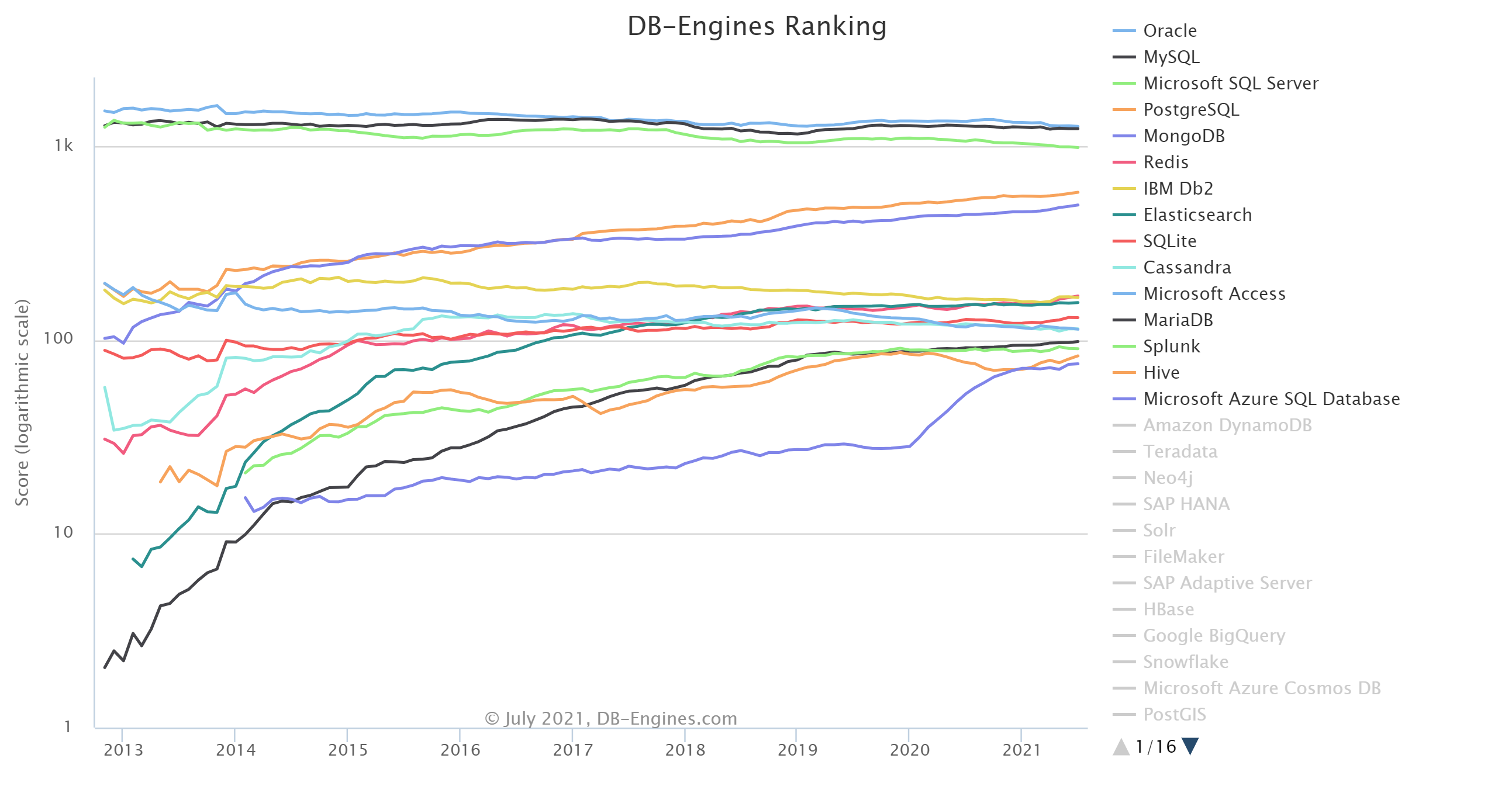}}
	\caption{Trend of Top 20  DBMS \cite{trendDBEngines2021}}
	\label{tcomplete}
\end{figure}

Figure \ref{tcomplete} presents popularity trend of top 20 database management systems since 2013. A close competition can be seen between Oracle, MySQL, and Microsoft SQL Server from the beginning of 2013. However, SQL Server has been on slight decrease in popularity and the gap is getting wider. Popularity of PostgreSQL is on constant rise and getting closer to SQL Server. MongoDB has become one of the fasted growing document store system which shows the interest from the developers community in model different than relational.

\section{Analysis}
\label{analysisSec}
This section breaks down the popularity of database management systems by model and provides a brief analysis. 

\subsection{Number of systems per category}
\begin{figure}[htbp]
	\centerline{\includegraphics [width=0.47\textwidth]{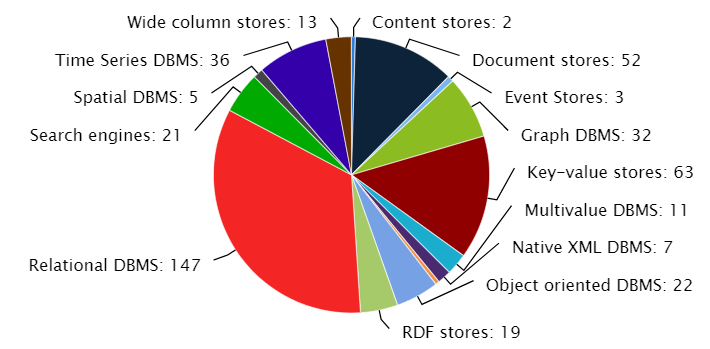}}
	\caption{Number of Systems Per Category \cite{analysisDBEngines2021}}
	\label{anumber}
\end{figure}

Figure \ref{anumber} presents number of systems of per category. There are total 373 database management systems of which 40\% are relational systems followed by 17\% of Key-value stores. Relational DBMS, Document stores, and key-value stores comprise of total 262 systems (70\%) out of 373. 

\subsection{Ranking score per category}
\begin{figure}[htbp]
	\centerline{\includegraphics [width=0.47\textwidth]{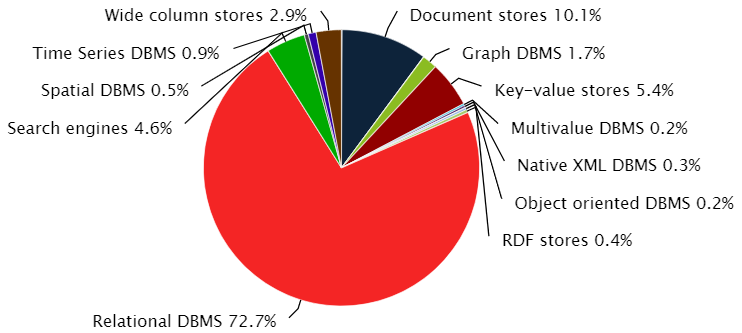}}
	\caption{Ranking Score Per Category \cite{analysisDBEngines2021}}
	\label{apercent}
\end{figure}

Figure \ref{apercent} shows ranking score per category in percentage for July, 2021. Popularity score of all individual systems for each category is used to calculate total percentage. As Table \ref{rcomplete} showed that Oracle, MySQL, and MS SQL Server makes highest score for all relational systems therefore this is also depicted in Figure \ref{apercent} where relational DBMS have 72.7\% of total popularity. Document stores are other set of systems which are getting popular with MongoDB being most popular of them all.

\subsection{Popularity Changes}

The following set of charts show the historical trend of the categories' popularity. In the ranking of each month the best three systems per category are chosen and the average of their ranking scores is calculated. In order to allow comparisons, the initial value is normalized to 100. 
\subsubsection{Complete Popularity Trend}
\begin{figure}[htbp]
	\centerline{\includegraphics [width=0.47\textwidth]{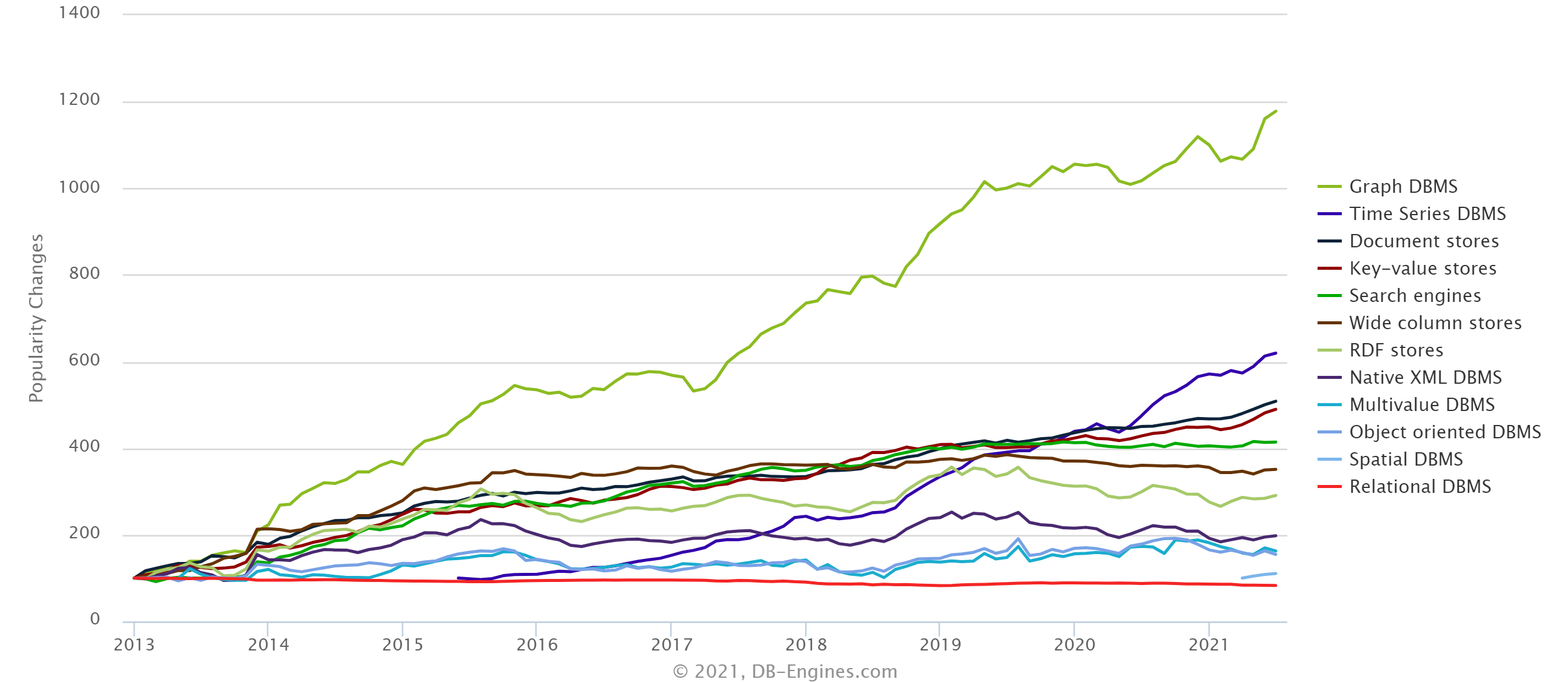}}
	\caption{Complete Popularity Trend \cite{analysisDBEngines2021}}
	\label{apopularity}
\end{figure}

The chart in \ref{apopularity} shows that Graph DBMS has been on constant rise in popularity because of their main use in social networking websites such as Facebook, Twitter, and TikTok. Time Series systems are another type of DBMSs which have gain popularity in recent times due to advancements in IOT technology.
 
\subsubsection{Trend for last 24 months}
\begin{figure}[htbp]
	\centerline{\includegraphics [width=0.47\textwidth]{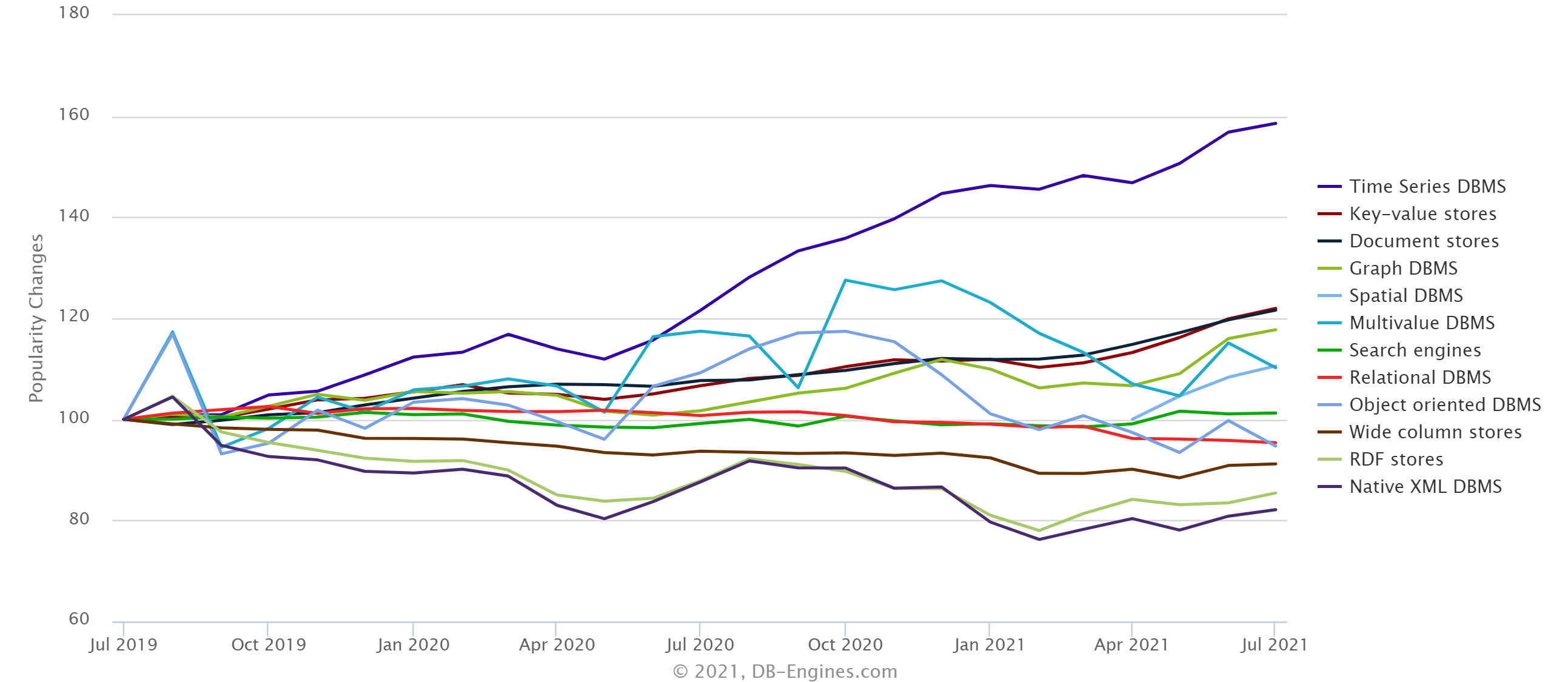}}
	\caption{Trend for last 24 months \cite{analysisDBEngines2021}}
	\label{atrend24}
\end{figure}

This chart \ref{atrend24} shows change in popularity trend for last 24 months. As mentioned earlier, Time Series DBMS has seen rise in popularity in recent times due to their use in Internet of Things. Relational systems on the other hand are at the same level of popularity as in July 2019. However, DBMS like Native XML and RDF stores are on decline.

\section{Conclusion and Future Work}
\label{conclusion}
Databases are considered to be integral part of modern information systems. Almost every web or mobile application uses some kind of database. Database management systems are considered to be a crucial element from both business and technological standpoint. This paper divided different types of database management systems into two main categories i.e. relational and non-relational and several sub categories. Ranking of various sub categories for the month of July, 2021 are presented in the form of popularity score calculated and managed by DB-Engines. Popularity trend for each category is also presented to look at the change in popularity since 2013. Complete ranking and trend of top 20 systems has shown that relational models are still most popular systems comprising of 12 systems in the top 20 with top 4 all being relational systems. Relational DBMS also make the most number of systems and highest percentage in popularity. Oracle and MySQL being two most popular systems. However, recent trends have shown DBMSs like Time Series and Document Store getting more and more popular with their wide use in IOT technology and BigData, respectively. 

Future work can include deep analysis of each category with more focus on last 24 months. A comparative analysis for Document Store systems versus Relational systems can present with the results on whether systems like MongoDB and Google Firebase can challenge hugely popular systems like Oracle and MySQL. It will also be an interesting to divide the systems into commercial and open-source categories and then looking at popularity ranking and trends.

\bibliographystyle{IEEEtran}

% The bibliography should be embedded for final submission.

\bibliography{oodbib}

\end{document}